\documentclass[journal]{IEEEtran}

\usepackage{comment}
\usepackage{soul}
\usepackage{color}
\usepackage[usenames,dvipsnames]{xcolor}

\usepackage{cite}
\usepackage{url}

\ifCLASSINFOpdf
\else
\fi
\usepackage{tikz}
\usetikzlibrary{arrows, positioning, shapes.geometric, fit, calc,backgrounds,shapes.arrows}
\usetikzlibrary{arrows.meta}

\usepackage{amsmath}
\usepackage{amsfonts}
\usepackage[short]{optidef}
\usepackage{algorithm}
\usepackage{algpseudocode}
\usepackage{ bbold }
\usepackage{amssymb}

\ifCLASSOPTIONcompsoc
  \usepackage[caption=false,font=normalsize,labelfont=sf,textfont=sf]{subfig}
\else
  \usepackage[caption=false,font=footnotesize]{subfig}
\fi

\usepackage{flushend}

\usepackage{bm}
\usepackage{hyperref}
\usepackage[capitalise]{cleveref}
\Crefname{equation}{}{}

\begin{document}
%
\title{Auction-Based vs Continuous Clearing in Local Flexibility Markets with Block Bids}

\author{
\IEEEauthorblockN{Alicia Alarcón Cobacho, Eléa Prat, Daniel Vázquez Pombo, Spyros Chatzivasileiadis}
\IEEEauthorblockA{Department of Electrical Engineering, Technical University of Denmark, Kongens Lyngby, Denmark}
a.alarcoba@gmail.com, \{emapr, dvapo, spchatz\}@elektro.dtu.dk
\thanks{This work is supported by the FLEXGRID project, funded by the European Commission Horizon 2020 program, Grant Agreement No. 863876.}
}


\maketitle

\begin{abstract} 
Flexibility markets can be introduced as a tool for the distribution system operator (DSO) to avoid high costs and public opposition against new network investments.
Continuous flexibility markets have the advantage of allowing more liquidity, which can be critical in the earlier stages of such markets, and can be operated closer to real-time, thereby enabling a better use of the latest forecasts; but, by design, they also result to a lower social welfare compared to auction-based markets. This paper has two main contributions. First, it introduces a continuous local flexibility market which includes both network constraints and asymmetric block bids. Second, it proposes an algorithm that can accurately determine the upper and lower bound of the social welfare loss compared with an auction-based clearing model.

\end{abstract}

\begin{IEEEkeywords}
asymmetric block bids, auction-based clearing, continuous clearing, local flexibility markets
\end{IEEEkeywords}

\section{Introduction \& Motivation}

Due to the increasing penetration of distributed energy resources (DERs), distribution networks are expected to be operated closer to their capacity limits. Wholesale markets, however, generally do not consider distribution network constraints, which can often lead to limit violations at distribution level (e.g. line limits, voltage limits, etc.). Until recently, a more costly dispatch or grid reinforcement would have been necessary to avoid such violations. This paper explores a continuous local energy market model that can harness the flexibility available in the distribution system and defer costly investments in the distribution grid.

Indeed, the recent report \cite{ENTSOE} published by ENTSO-E together with the major European associations of DSOs emphasises the requirement for grid flexibility procurement. 
Ref.~\cite{Schitte} recognises flexibility markets as the tool needed in Europe to make a more efficient use of the existing distribution grid. Local flexibility markets (LFMs) are introduced to promote flexibility trade in limited areas such as communities or small towns in Ref.~\cite{Jin2020}. The authors analyse the concepts and models proposed so far for LFMs, defining them as platforms that connect actors requiring flexibility with actors offering it. In this context, flexibility is identified as a controlled power variation that can be performed at a localized point in the network, with a given duration and at a specific time. 

Ref.~\cite{Schitte} analyses four pioneering European projects that implement flexibility markets: Pico Flex, Enera, GOPACS and NODES. While Enera, GOPACS and NODES use continuous trading, Pico Flex, together with the Danish LFM project EcoGrid 2.0 \cite{EcoGrid} consider auction-based trading. In Ref.~\cite{FuturIntraday} the implications of using both mechanisms when designing future electricity intraday markets are discussed. With auction-based clearing, the bids accepted in the market are the ones that lead to the highest social welfare, whereas with continuous clearing bids are matched as they enter the market. The authors state that even if the continuous clearing results in a sub-optimal social welfare, it can allow for more trades. The same conclusion is drawn in Ref.~\cite{Comparison}, where continuous and auction-based energy markets are compared analytically, having as benchmark a discrete double auction. The authors also provide an upper bound on the sub-optimality of continuous clearing compared to auction-based clearing. In contrast to what we present in this paper, though, the compared models do not include network constraints or block bids, which would affect the outcome of the comparison.

Network constraints have been included in Ref.~\cite{cont}, which presents probably the first formulation of a continuous LFM considering network constraints. This market trades flexibility as reserve capacity following the first-come, first-served principle to match flexibility requests and offers and uses the pay-as-bid pricing rule. Including a network check when clearing the market is critical to ensure that the trades do not lead to any limits violations.

When it comes to block bids, both Ref.~\cite{DSOcontract}~and~\cite{CongestionManagement} account for the preferences of the flexibility providers through asymmetric block bids, which implies including integer variables in their models. These bids consist of several single offers linked to each other, submitted at the same time and location but with different direction, quantities and prices, and referred to different time targets. Papers \cite{Thermal-electric,Supermarket,FlexStrategy} present different approaches to provide flexibility from the demand side through this type of bids. Thermal electric loads such as building heating and cooling, water heating and refrigeration are ideal candidates for load shifting, as they can vary their energy consumption without compromising their purpose \cite{Thermal-electric}. Ref~\cite{Thermal-electric} models the rebound effect attached to them, designing block bids that can be only fully accepted or fully rejected, i.e All-or-Nothing condition (AoN). 
Ref.~\cite{Supermarket} builds the block bids needed by a supermarket refrigerator to supply flexibility, analysing its demand response capability. Ref.~\cite{FlexStrategy} develops an offering strategy for a flexibility aggregator to participate in a balancing market using asymmetric block bids.

Extending our previous work in Ref.~\cite{cont}, this paper has two main contributions: 
\begin{itemize}
    \item We introduce a continuous local flexibility market which explicitly includes both network constraints and block bids. To our knowledge, this is the first formulation of continuous energy markets that considers both network constraints and block bids.
    \item We introduce a method that can determine both the \emph{upper bound} and the \emph{lower bound} on the suboptimality of such continuous markets compared with their auction-based counterparts.
\end{itemize}

The rest of this paper is organised as follows. In Section \ref{sec:LFM} we describe the designed local flexibility market (LFM). We first present the market framework in which the two clearing mechanisms are set up and then we explain the formulation and performance of the continuous and auction-based models developed. The case study used to illustrate the operation of the LFM and the corresponding results for different configurations are given in Section \ref{sec:case}. The sub-optimality gap is further studied in Section \ref{sec:subopt}, where we propose a method to obtain the highest and lowest social welfare using continuous clearing. In Section \ref{sec:discussion}, we discuss the models designed and the clearing mechanisms compared. Finally, Section \ref{sec:ccl} presents the conclusions and directions for future work.

\section{Local Flexibility Market Design}
\label{sec:LFM}

Flexibility can locally be traded either as reserve or as energy. In this paper, we develop Local Flexibility Market (LFM) clearing algorithms in which flexibility is traded in the form of energy. Future work will extend this framework to reserve markets as well.
The platform connects actors who require flexibility (i.e. DSO) with actors who offer it (i.e. prosumers and aggregators), matching flexibility requests with flexibility offers. The flexibility market operator is an external agent, whom we assume the DSO provides with perfect information about the network topology and limits and the system setpoint.
When clearing the market, a DC power flow algorithm checks whether the power grid is technically able to handle the expected energy transfers without causing or aggravating congestions, similar to the auction-based market counterparts.

The LFM is multiperiod, to account for block bids that model the preferences of certain flexibility providers. Block bids consist of several single offers linked to each other, one for each time period covered by the block bid. 

Only offers can be submitted as block bids since they are mainly introduced to encourage suppliers' participation.
While single bids can be partially matched (i.e. the market can accept part of the offered energy), block bids can only be fully accepted or rejected, as proposed in Ref.~\cite{CongestionManagement}. This condition and the asymmetric form of the block bids are both due to the technical requirements associated with the rebound effect.

Within this framework, two different models are presented: one with continuous clearing, and the other with auction-based clearing which will be used as a benchmark. The main difference is that a continuous market clears as soon as there is a match between bids, whereas an auction-based market clears once every time period considering all the bids submitted for that period. 

\subsection{Continuous Clearing Model}

The market clearing follows the first-come, first-served principle, which means that the older bid sets the clearing price and has priority for the same price. Every time a bid enters the market, we examine all the previous requests/offers standing in the corresponding order book. A match can occur if a request and an offer have the same direction (upward or downward) and time target, and if the request price is higher or equal to the offer price. The unmatched requests are stored by descending price and the unmatched offers are ordered by ascending price in their respective order books. This ensures that the incoming bid is matched at the best available price, so the social welfare is the highest possible for each match.  

Once a match is found in terms of direction, time target and price and if only single bids are matched, the model calculates the \textit{Quantity\_max} to be exchanged through the algorithm introduced in Ref.~\cite{cont}. It performs a DC power flow analysis using PTDFs factors that link power injections with line flows. The system setpoint is updated after every match.

Similar to the single bids, block bids should be matched with the best available requests. However, due to the AoN condition, we have to ensure that all the single offers included in the block bid can be fully matched before setting a match. To prioritise seniority, older requests should be temporarily assigned to the block bid and stored until its complete match is possible. To be fair across all incoming offers, however,, the requests assigned to a block bid should still be available in their order book, in case a new matching offer appears before the match with the block bid becomes effective. Then, if a temporary match with a block bid is partially or completely cancelled, the requests order book should be immediately revisited looking for a new match for the remaining offer. Furthermore, every match that occurs while the block bid is not fully matched changes the line flows and can, thus, technically limit the match of the block bid. Therefore, network constraints should be constantly checked to guarantee that the temporary matches with the block bid are still feasible.

To avoid this tedious process, we store all the possible matches with the single offers involved in the block bid as candidates, until there are enough candidate requests to fully match it. After meeting this condition, it may happen that some parts of the block bid have several candidates to match with. To determine which one(s) to choose we run an optimisation problem, considering all the possible matches with the single offer concerned. This way, we can select the request(s) that lead to the highest social welfare respecting network constraints. The optimisation problem is a linear program (LP) formally defined as follows:
\begin{subequations}
  \begingroup
  \allowdisplaybreaks
  \begin{align}
    & \underset{\mathbf{x}}{\min} && 
	 \lambda _{b,t}^{\text{U}}  P_{b,t}^{\text{U}}+\lambda _{b,t}^{\text{D}}  P_{b,t}^{\text{D}} -\sum_{r \in \mathcal{R}_{b}} \left ( \lambda _{r,t}^{\text{U}}  P_{r,t}^{\text{U}}+\lambda _{r,t}^{\text{D}} 
    P_{r,t}^{\text{D}}\right)
    \label{eq:01Obj}\\
	& \text{s.t.} 
	&& P_{n,t}^{\text{S}} + P_{b,t}^{\text{U}}-P_{b,t}^{\text{D}} 
	- \sum_{r \in \mathcal{R}_{n}}\left ( P_{r,t}^{\text{U}} -P_{r,t}^{\text{D}} \right ) \notag\\
	& && - \sum_{m \in \Omega_{n}}\left ( b_{n,m}\left ( \delta_{n,t} - \delta_{m,t}\right ) \right ) = 0 \quad n = n_{b} \label{eq:02NodBal1}\\
	& && P_{n,t}^{\text{S}} - \sum_{r \in \mathcal{R}_{n}}\left ( P_{r,t}^{\text{U}} -P_{r,t}^{\text{D}} \right ) \notag\\
	& && - \sum_{m \in \Omega_{n}}\left ( b_{n,m}\left ( \delta_{n,t} - \delta_{m,t}\right ) \right ) = 0 \quad \forall{n} \in \mathcal{N}, n \neq n_{b}  \label{eq:02NodBal2}\\
    & && -P_{n,m}^{\text{lim}}\leq b_{n,m} \left ( \delta_{n} - \delta_{m}\right )\leq P_{n,m}^{\text{lim}} \quad \forall n \in \mathcal{N}, m \in \Omega_{n} \label{eq:O3LineLim} \\
    & && 0\leq P_{r}^{\text{U}}\leq P_{r}^{\text{Umax}} \quad \forall {r} \in \mathcal{R}\label{eq:O4RUmax} \\
    & && 0\leq P_{r}^{\text{D}}\leq P_{r}^{\text{Dmax}} \quad \forall {r} \in \mathcal{R}\label{eq:O4RDmax} \\
    & && \delta_{\text{ref}} = 0 \label{eq:O9ref}
  \end{align}
  \endgroup
  \label{eq:problemO.0}
\end{subequations}
\normalsize
where $\mathbf{x}=\{P_{r,t}^{\text{U}},P_{r,t}^{\text{D}},\delta_{n,t}\}$.

The objective function \eqref{eq:01Obj} seeks to minimise total costs, given the cost of the offer in question, and the sum of the costs of the requests that constitute the complete match. The decision variables are $P_{r}$, the energy fulfilling request $r$ and $\delta_{n,t}$ the voltage angle at node $n$. The price and the quantity bid for the single offer $b$ contained in the block bid $k$ are given in $\lambda _{b,t}$ and $P_{b,t}$. The information about the requests are $\lambda _{r,t}$, the price of the request $r$ and $\mathcal{R}_{b}$, the set of candidate requests to match with $b$. The superscript $U$ stands for upward and $D$ for downward.

The nodal balance at the node where the block bid is located, $n_{b}$, is represented in \eqref{eq:02NodBal1}, whereas \eqref{eq:02NodBal2} applies for the rest of the nodes, contained in the set $\mathcal{N}$.
$P_{n,t}^{\text{S}}$ is the initial setpoint of node $n$ at time period $t$ and $\mathcal{R}_{n}$ is the set of requests located at node $n$. The last term of both equations refers to the energy flows from or to the node, with $\Omega _{n}$ being the sets of nodes connected to $n$. These flows are calculated using the susceptance of the line, $b_{n,m}$, and the voltage angle difference of the connecting nodes $n$ and $m$. 

Constraint \eqref{eq:O3LineLim} guarantees that the energy flow through each line respects its capacity limit in both directions, $-P_{n,m}^{\text{lim}}$ and $P_{n,m}^{\text{lim}}$. Constraint \eqref{eq:O4RUmax} makes the amount of energy traded for each request positive and equal or lower than the quantity bid, $P_{r}^{\text{max}}$. Finally, the last constraint \eqref{eq:O9ref} defines that the voltage angle $\delta$ at the reference node is always $0$.

As the optimisation problem is solved separately for each single offer composing the block bid,  subscript $t$ corresponds to the time target for the given single offer $b$. Since $b$ has only one direction, upwards ($U$) or downwards ($D$), the terms in the opposite direction are disregarded along the whole problem.

The entire process carried out to match block bids is described in Algorithm~\ref{alg:BBmatching}. In case the possible match is between single bids, or there is only one candidate request for the single offer of a block bid, we use the PTDF method to check network constraints, similar to \cite{cont}, rather than solving the optimisation problem \eqref{eq:problemO.0}, in order to reduce computational complexity.
\begin{algorithm}
\caption{Block Bids Matching}\label{alg:BBmatching}
\begin{algorithmic}
\small
\For{each \textit{offer} of the \textit{block bid}}
    \If {there is just one possible match}
        \State Calculate \textit{Quantity\_max} following Ref.~\cite{cont}
        \If {\textit{Quantity\_max} = \textit{Quantity\_bid}}
            \State Save potential match and move to the next \textit{offer}
        \Else {}
            \If {\textit{Quantity\_max} $<$ \textit{Quantity\_bid}}
                \State Store the \textit{request} as candidate match and exit
            \EndIf
        \EndIf    
    \Else {}
        \If {there are multiple possible matches}
            \State Solve (\ref{eq:problemO.0}) to determine the best match
            \If {(\ref{eq:problemO.0}) is feasible}
                \State Save potential match and move to the next \textit{offer}
            \Else{}
                \State Store the \textit{requests} as candidate match and exit
            \EndIf
        \EndIf
    \EndIf
\EndFor
\If {there is a potential match for each \textit{offer} of the \textit{block bid}}
    \State Set a match
    \State Update the Setpoint
\EndIf
\normalsize
\end{algorithmic}
\end{algorithm}

As a result of this continuous clearing model, single bids are matched with the best available option at the time of their submission, and block bids are matched with the set of the best options available at the moment when the full match of the block bid is possible. For all matches, the proposed market clearing guarantees that network constraints will be satisfied.

\subsection{Auction-Based Clearing Model}

The auction-based clearing, which will serve as a benchmark for the proposed continuous clearing algorithm, is built as a mixed integer linear program (MILP), since it accounts for block bids using binary variables as proposed in Ref.~\cite{CongestionManagement}. These variables represent the acceptance ratio of the block bids, with 1 standing for acceptance and 0 for rejection. The aim of the auction-based configuration is to achieve the maximum social welfare for the whole market time horizon. Due to the AoN acceptance condition of the block bids, all the time periods are cleared at once.
The problem is formally defined as follows:
\begin{subequations}
  \begingroup
  \allowdisplaybreaks
  \begin{align}
    & \underset{\mathbf{x}}{\text{min.}} && 
	 \sum_{t \in \mathcal{T}} \Bigg[ \sum_{o \in \mathcal{O}}\left( \lambda _{o,t}^{\text{U}}  P_{o,t}^{\text{U}}+\lambda _{o,t}^{\text{D}}  P_{o,t}^{\text{D}} \right) \notag\\
	 & &&+\sum_{k \in \mathcal{K}}\left(AR_{k} \sum_{b \in \mathcal{B}_{k}} \left( \lambda _{b,t}^{\text{U}}  P_{b,t}^{\text{U}}+\lambda _{b,t}^{\text{D}}  P_{b,t}^{\text{D}} \right)\right)
     \notag\\
    & &&-\sum_{r \in \mathcal{R}} \left ( \lambda _{r,t}^{\text{U}}  P_{r,t}^{\text{U}}+\lambda _{r,t}^{\text{D}} 
    P_{r,t}^{\text{D}}\right) \Bigg]  \label{eq:O1.1Obj}\\
	& \text{s.t.} 
	&&  P_{n,t}^{\text{S}} - \sum_{r \in \mathcal{R}_{n}}\left ( P_{r,t}^{\text{U}} -P_{r,t}^{\text{D}} \right ) + \sum_{o\epsilon \mathcal{O}_{n}}\left ( P_{o,t}^{\text{U}}-P_{o,t}^{\text{D}} \right )  \notag\\
	& && + \sum_{k \in \mathcal{K}_{n}}\left (AR_{k} \sum_{b \in \mathcal{B}_{k}} \left( P_{b,t}^{\text{U}}+  P_{b,t}^{\text{D}} \right)\right )  \notag\\
	& && - \sum_{m \in \Omega_{n}}\left ( b_{n,m}\left ( \delta_{n,t} - \delta_{m,t}\right ) \right ) = 0, \quad \forall{n} \in \mathcal{N},\forall {t} \in \mathcal{T} \label{eq:02.1NodBal}\\
	& && \sum_{o \in \mathcal{O}}\ P_{o,t}^{\text{U}} + \sum_{k \in \mathcal{K}} AR_{k} P_{k,t}^{\text{U}} - \sum_{r \in\mathcal{R}} P_{r,t}^{\text{U}} = 0, \quad \forall {t} \in \mathcal{T}\label{eq:02.1up}\\
	& && \sum_{o \in \mathcal{O}}\ P_{o,t}^{\text{D}} + \sum_{k \in \mathcal{K}} AR_{k} P_{k,t}^{\text{D}} - \sum_{r \in \mathcal{R}} P_{r,t}^{\text{D}} = 0 \quad \forall {t} \in \mathcal{T}\label{eq:02.1down}\\
    & && -P_{n,m}^{\text{lim}}\leq b_{n,m} \left ( \delta_{n,t} - \delta_{m,t}\right )\leq P_{n,m}^{\text{lim}} \notag\\
	& &&\quad \forall n \in \mathcal{N}, m \in \Omega_{n},\forall {t} \in \mathcal{T}\label{eq:O3.1LineLim}\\ 
    & && 0\leq P_{r,t}^{\text{U}}\leq P_{r,t}^{\text{Umax}} \quad \forall {r} \in \mathcal{R},\forall {t} \in \mathcal{T}\label{eq:O4.1RUmax}\\
    & && 0\leq P_{r,t}^{\text{D}}\leq P_{r,t}^{\text{Dmax}} \quad \forall {r} \in \mathcal{R},\forall {t} \in \mathcal{T}\label{eq:O5RDmax}\\
    & && 0\leq P_{o,t}^{\text{U}}\leq P_{o,t}^{\text{Umax}} \quad \forall {o} \in \mathcal{O},\forall {t} \in \mathcal{T}\label{eq:O6OUmax}\\
    & && 0\leq P_{o,t}^{\text{D}}\leq P_{o,t}^{\text{Dmax}} \quad \forall {o} \in \mathcal{O},\forall {t} \in \mathcal{T}\label{eq:O7ODmax}\\
    & && AR_{k}\: \epsilon \left\{0,1 \right\} \quad \forall {k} \in \mathcal{K}\label{eq:O8Binary}\\
    & && \delta_{\text{ref},t} = 0 \quad \forall {t} \in \mathcal{T}
     \label{eq:O9.1ref}
  \end{align}
  \endgroup
  \label{eq:problemO}
\end{subequations}
\normalsize
The objective function \eqref{eq:O1.1Obj} minimises the cost of trading flexibility with $\mathbf{x}=\{P_{r,t}^{\text{U}},P_{r,t}^{\text{D}},P_{o,t}^{\text{U}},P_{o,t}^{\text{D}},\delta_{n,t}, AR_{k}\}$. It considers all the single offers $\mathcal{O}$, block offers $\mathcal{K}$ and requests $\mathcal{R}$ submitted for all the time periods $\mathcal{T}$. The first and the third term refer to the cost of single offers and requests. The second term is the sum of the cost of all single offers contained in a block bid $\mathcal{B}_{k}$, multiplied by the acceptance ratio $AR_{k}$ of the block bid $k$, defined in \eqref{eq:O8Binary}. 

The first constraint \eqref{eq:02.1NodBal} is the nodal balance, quite similar to \eqref{eq:02NodBal1} but adding a term for the block bids and considering all the offers located at each node $\mathcal{O}_{n}$ and $\mathcal{K}_{n}$. 
Constraints \eqref{eq:O3.1LineLim} and \eqref{eq:O9.1ref} stand for the power flow limits and the reference voltage angle, exactly in the same way as in \eqref{eq:O3LineLim} and \eqref{eq:O9ref}. The energy accepted per bid is limited in \eqref{eq:O4.1RUmax} - \eqref{eq:O7ODmax}, where $P^\text{max}$ is the total quantity bid. Solving the optimisation problem, we determine which bids to accept and for which quantity in order to achieve the highest social welfare respecting network constraints.

\section{Case Study}
\label{sec:case}

\subsection{Case Description}

To compare both clearing mechanisms we design a case study where we apply the market models presented. We operate the market for 24 hours in the 33-bus radial distribution grid introduced in \cite{Grid}, considering consumption acting as flexible loads and including small-scale renewable generation as DERs, and allow for flexibility offers. The initial setpoint for 24 hours is obtained combining an average daily load profile for a distribution grid with typical generation daily profiles for photovoltaic plants and wind farms. We start from an infeasible setpoint, which violates line limits, aiming to make it feasible by trading system flexibility.

We assume that we have perfect information from the DSO in terms of topology and technical specifications of the network, as well as the power injection setpoints per node and time period. We assume that the power injection setpoint is defined after the clearing of the wholesale market. Through a DC-OPF analysis we determine the flexibility requests needed by the DSO to operate the system in a feasible way. Upward requests represent the load shedding and downward requests the curtailment required in case the DSO does not have access to trade flexibility. Flexibility offers are created randomly to simulate that they can be submitted whenever and wherever. 

The prices of flexibility requests and offers are randomly generated within certain intervals. We consider that the DSO sets them according to the damage caused when capping generation and shedding loads. As the DSO usually wants to avoid load shedding at any cost, the prices of upward requests, [0.250-0.300]€/kWh, are quite high compared to downward requests, [0.035-0.045]€/kWh, to guarantee their acceptance. As a starting point, we consider that the price of flexibility offers is close to the wholesale market price in both directions, [0.030-0.040]€/kWh. Block bids price is assumed to be a bit lower to promote their acceptance, since it is more difficult for them due to their AoN condition, [0.020-0.035]€/kWh.

The data used is available online, as well as the implemented models \cite{github}. Both market configurations are developed in Pyomo, a Python-based environment targeting optimisation problems \cite{pyomo} and solved using Gurobi \cite{gurobi}.

\subsection{Simulation and Results}
\begin{figure}[t]
    \centering
    \includegraphics[width=0.95\linewidth]{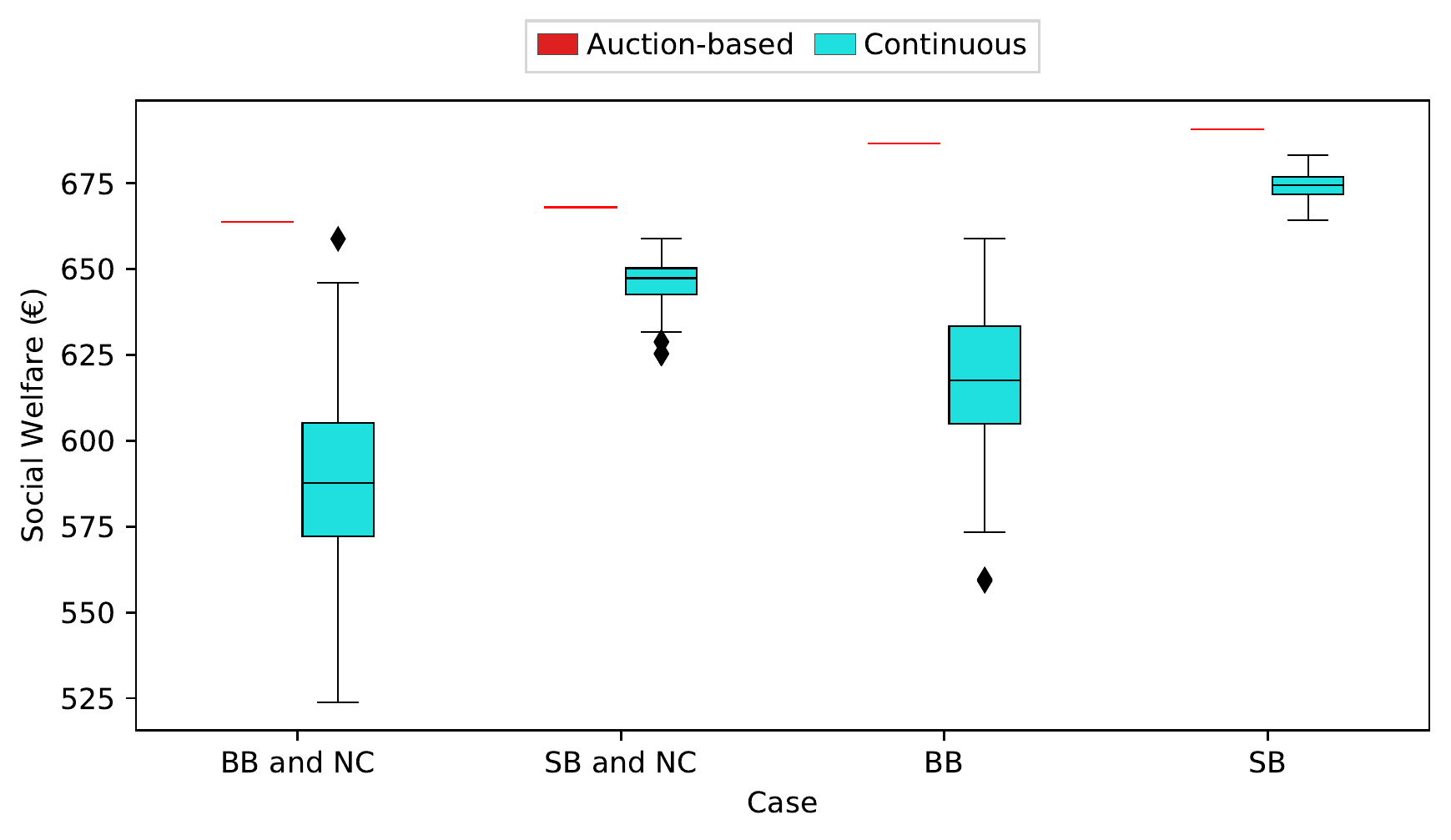}
    \caption{Social welfare for the four cases using continuous and auction-based clearing.}
    \label{fig:SW}
\end{figure}

\begin{figure}[t]
    \centering
    \includegraphics[width=0.95\linewidth]{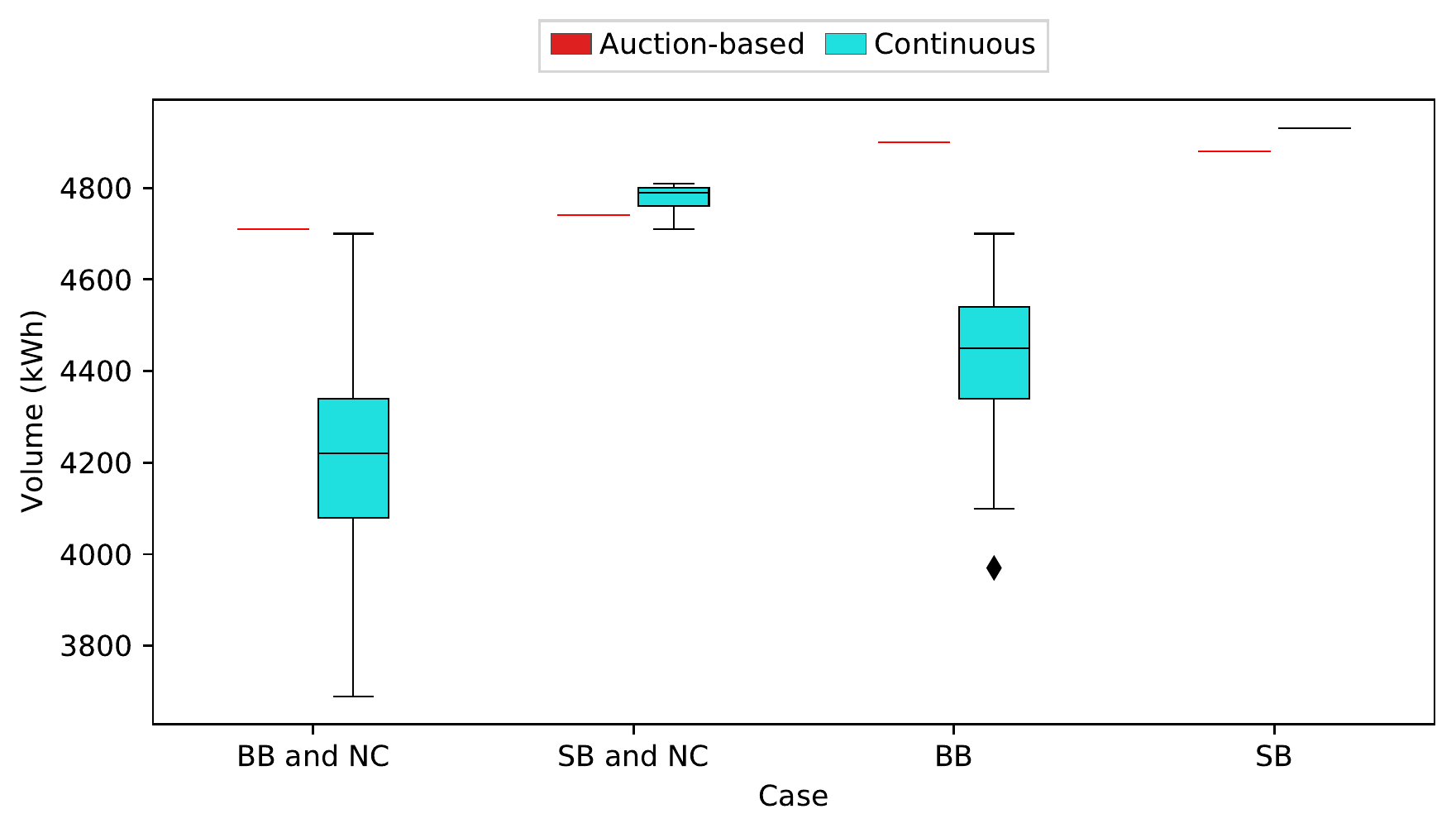}
    \caption{Energy volume traded for the four cases using continuous and auction-based clearing.}
    \label{fig:Vol}
\end{figure}

The performance of the models is evaluated focusing on the social welfare and the volume of energy traded, both for the whole market horizon (i.e. 24 hours). To analyse the differences between the clearing mechanisms, we define four test cases, containing the same bids but varying their form:
\begin{itemize}
    \item \textit{BB and NC}: Single and block bids (BB) with network constraints (NC).
    \item \textit{SB and NC}: Only single bids (SB) with network constraints, i.e. block bids are treated as independent single bids.
    \item \textit{BB}: Single and block bids without network constraints.
    \item \textit{SB}: Only single bids without network constraints.
\end{itemize}
Being aware of the continuous clearing dependence on the arrival order of the bids, we generate a set of 100 scenarios with the same bids, but random arrival sequences. Fig.~\ref{fig:SW} and Fig.~\ref{fig:Vol} show the total social welfare and energy volume traded for all the scenarios when operating both models. The auction-based market results are constant because they are not affected by the arrival order of the bids.

Comparing the results of the auction-based clearing in Fig.~\ref{fig:SW}, we observe that the integration of network constraints has more impact on the social welfare than the introduction of block bids. With the continuous clearing, however, it is generally the other way around, i.e. block bids affect more the social welfare than network constraints. Considering only single bids, the social welfare in the continuous market can be closer to the auction-based, whereas when including block-bids the social welfare difference increases.
As an insight we get that block bids might be more difficult to handle in the continuous market, which results to larger differences in social welfare between the clearing mechanisms.

Regarding the energy volume traded displayed in Fig.~\ref{fig:Vol}, we notice that with single bids, continuous clearing usually leads to more liquidity. Without considering block bids nor network constraints, all the scenarios for the continuous clearing model result in the same volume, which is the maximum according to the bids submitted. The difference we observe in the volume between the continuous clearing and the auction-based clearing in this case lies in the acceptance of some bids that do not add to the social welfare (social welfare of zero). Our auction-based model maximises social welfare, so it does not accept bids that bring no social welfare. In contrast, our continuous model does consider the match between bids with the same price, i.e. zero social welfare. For this reason, there are more transactions and hence market liquidity increases.

Table~\ref{tab:comparison} presents the continuous market performance relative to the auction-based market outcome in terms of social welfare and volume. The difference between them is displayed as percentages of the auction-based result, since it is considered as benchmark.

\begin{table}[t]
\caption{Continuous vs auction-based clearing (values in \% relative to the auction-based Social Welfare and Volume).}
\resizebox{0.5\textwidth}{!}{
\label{tab:comparison}
\begin{tabular}{l|ccc|ccc|}
\cline{2-7}
\multicolumn{1}{c|}{\textbf{}} & \multicolumn{3}{c|}{\textbf{Social welfare}} & \multicolumn{3}{c|}{\textbf{Volume}} \\ \cline{2-7} 
\multicolumn{1}{c|}{}          & Average        & Max          & Min          & Average    & Max        & Min        \\ \hline
\multicolumn{1}{|l|}{BB and NC}   & 88.6\%         & 99.2\%       & 78.9\%       & 89.5\%     & 99.9\%     & 78.3\%     \\
\multicolumn{1}{|l|}{SB and NC}   & 96.7\%         & 98.7\%       & 93.6\%       & 100.9\%    & 101.5\%    & 99.4\%     \\
\multicolumn{1}{|l|}{BB}   & 89.9\%         & 96.0\%       & 81.4\%       & 90.6\%     & 95.9\%     & 81.0\%     \\
\multicolumn{1}{|l|}{SB}   & 97.6\%         & 98.9\%       & 96.2\%       & 101.0\%    & 101.0\%    & 101.0\%    \\ \hline
\end{tabular}
}
\end{table}

The results of this case study show that including block bids, the sub-optimality gap in social welfare between continuous and auction-based clearing is on average slightly higher than 10\%, whereas without them is around 3\%. The difference between considering network constraints or not is generally around 1\%. With single bids and without network constraints the volume can increase by 1\% with continuous clearing, and up to 1.5\% considering network constraints. However, with block bids it is reduced by around 10\%. In general, we can conclude that when we only have single bids, the bids arrival order has less influence on the continuous market results as when we introduce block bids. When we have block bids, the sub-optimality gap in terms of social welfare and volume can be between 1\% and 21\% with network constraints and 4\% and 19\% without them. We therefore observe that the arrival order of the bids, which is initially unpredictable, has a considerable influence on the outcome of the continuous market, and hence it widens the gap between the clearing mechanisms. The next section develops an approach to determine the upper and lower bound of this sub-optimality gap between the continuous and the auction-based markets, by determining the worst and best arrival sequence of the bids. 

\section{Study on the Sub-Optimality}
\label{sec:subopt}
In the previous case study, we generated random scenarios regarding the sequence of submission of the bids. Here, we want to analyse exactly how close and how far from the auction-based market the continuous market clearing can result in terms of social welfare. In, \cite{Comparison}, the best and worst arrival sequences are defined by ordering the offers by ascending and descending price, assuming that all requests are submitted previously. However, this does not apply for markets with network constraints and block bids. Indeed, network constraints limit the matching of the bids to prevent line congestions, and block bids disturb the best (and worst) arrival sequence, since all the offers involved are submitted at the same time. This fact together with their AoN acceptance condition also influences the matching. This section proposes an algorithm that can indeed define the best and worst arrival sequences, and thus determine the sub-optimality gap for a given set of bids, when considering both network constraints and block bids.

\subsection{Method}
Under the assumption that all requests are submitted first and standing in the order book, the worst (and best) sequence can be obtained by solving a multi-level optimisation problem, which reproduces the continuous market clearing, with the objective of minimising (or maximising) the resulting social welfare. It can be written as follows:

\small
\begin{subequations}
  \begingroup
  \allowdisplaybreaks
  \begin{align}
    & \underset{\mathbf{x}}{\min} &&
     \sum_{r\epsilon \mathcal{R}} \lambda _{r}  P_{r}^{\text{tot}} - \sum_{o\epsilon \mathcal{O}} \lambda _{o}  P_{o}^{\text{tot}} - \sum_{k\epsilon \mathcal{K}} AR_{k}^{\text{tot}} \left(\lambda_{k}^\text{U}  P_{k}^\text{U,max} - \lambda _{k}^\text{D}  P_{k}^\text{D,max} \right) \label{eq:UL_obj}\\
	& \text{s.t.} && C_i, \quad \forall{i} \in \mathcal{I} \label{eq:UL_LL}\\\
	& && P_{r}^{\text{tot}} = \sum_{i} P_{i,r}, \quad \forall r \in \mathcal{R} \label{eq:UL_Pr} \\
	& && P_{o}^{\text{tot}} = \sum_{i} P_{i,o}, \quad \forall o \in \mathcal{O} \label{eq:UL_Po} \\
	& && AR_{k}^{\text{tot}} = \sum_{i} AR_{i,k}, \quad \forall k \in \mathcal{K} \label{eq:UL_ARsum} \\
	& && AR_{k}^{\text{tot}} \in \left\{0,1 \right\} \quad \forall k \in \mathcal{K} \label{eq:UL_AR_bin}\\
	& && \sum_{b \in \mathcal{B}} s_{i,b} = 1 \quad \forall i \in \mathcal{I} \label{eq:UL_bin1} \\
	& && \sum_{i \in \mathcal{I}} s_{i,b} = 1 \quad \forall {b} \in \mathcal{B} \label{eq:UL_bin2} \\
	& &&  s_{i,b} \in  \left\{0,1 \right\} \quad \forall i \in \mathcal{I}, \forall s \in \mathcal{B} \label{eq:UL_bin}
  \end{align}
  \endgroup
  \label{eq:UL}
\end{subequations}
\normalsize
where $\mathbf{x}=\{P_{r}^\text{tot},P_{o}^\text{tot},AR_{k}^\text{tot}, s_{i,b}\}$.
Solving this problem returns one of the sequences in which a given set of offers submitted to the continuous market would result in the lowest social welfare (or highest if we use maximise instead of minimise). The value of the corresponding social welfare is available through the objective function \eqref{eq:UL_obj}, where $\lambda_r$, $\lambda_o$, $\lambda_k^\text{U}$ and $\lambda_k^\text{D}$ are the submitted prices of requests, single offers, upward and downward blocks of a block bid, and $P_{r}^\text{tot}$, $P_{o}^\text{tot}$, $P_{k}^{\text{U,max}}$ and $P_{k}^{\text{D,max}}$ are the associated quantity. For the block bids, those quantities are parameters and the all-or-nothing acceptance is decided through the variable $AR_{k}^\text{tot}$, which is defined as a binary in \eqref{eq:UL_AR_bin}. The set $\mathcal{R}$ gathers all requests, $\mathcal{O}$ all the single offers and $\mathcal{K}$ the block bids. These two are gathered in $\mathcal{B}=\mathcal{O}\cup \mathcal{K}$. The set $\mathcal{I}$ represents the different clearing rounds and has the same number of elements as $\mathcal{B}$: $|\mathcal{I}| = |\mathcal{B}|$. In \eqref{eq:UL_LL}, $C_i$ represent the round $i$ of the continuous clearing. The expression of the clearing will be later detailed. The auxiliary variables $P_{i,r}$, $P_{i,o}$ and $AR_{i,k}$ are defined to detail what is happening at each round of the market clearing. They sum up to the total quantities, as given in \eqref{eq:UL_Pr}, \eqref{eq:UL_Po} and \eqref{eq:UL_ARsum}. Finally, $s_{i,b}$ are introduced as binary variables \eqref{eq:UL_bin} to define the sequence for submitting the offers. When equal to 1, it means that offer $b$ is submitted in round $i$. The last constraints \eqref{eq:UL_bin1} and \eqref{eq:UL_bin2} ensure that only one bid is submitted at a time and that each bid is submitted only once.

The continuous matching $C_i$ can also be described by an optimization problem:

\small
\begin{subequations}
  \begingroup
  \allowdisplaybreaks
  \begin{align}
    & \underset{\mathbf{y}}{\max} &&
     \sum_{r \in \mathcal{R}} \lambda _{r}  P_{i,r} - \sum_{o \in \mathcal{O}} \lambda _{o}  P_{i,o}  \notag \\
     & && - \sum_{k \in \mathcal{K}} AR_{i,k} \left (\lambda _{k}^\text{U}  P_{k}^\text{U,max} + \lambda _{k}^\text{D}  P_{k}^\text{D,max}\right) \label{eq:LL_obj}\\
	& \text{s.t.} 
	&&  \sum_{o \in \mathcal{O}_{t}^\text{U}} P_{i,o} + \sum_{k \in \mathcal{K}_{t}} P_{i,k}^\text{U} = \sum_{r \in \mathcal{R}_{t}^\text{U}} P_{i,r}, \quad \forall{t} \in \mathcal{T} \label{eq:LL_sumU} \\
	& && \sum_{o \in \mathcal{O}_{t}^{\text{D}}} P_{i,o} + \sum_{k \in \mathcal{K}_{t}} P_{i,k}^\text{D} = \sum_{r \in \mathcal{R}_{t}^\text{D}} P_{i,r}, \quad \forall{t} \in \mathcal{T} \label{eq:LL_sumD}\\
	& && P_{n,t}^{\text{S}} + \sum_{j \in \mathcal{I}, j\leq i} \left ( \sum_{o \in \mathcal{O}_{n,t}^{U}} P_{j,o} - \sum_{o \in \mathcal{O}_{n,t}^{D}} P_{j,o}  \right. \notag\\
	& && \left.  + \sum_{k \in \mathcal{K}_{n,t}} (P_{j,k}^\text{U} - P_{j,k}^\text{D}) - \sum_{r \in \mathcal{R}_{n,t}^\text{U}} P_{j,r} + \sum_{r \in \mathcal{R}_{n,t}^\text{D}} P_{j,r} \right ) \notag\\
	& && - \sum_{m \in \Omega _{n}}b_{n,m}\left ( \delta_{i,n,t} - \delta_{i,m,t}\right ) = 0,  \quad \forall{n} \in \mathcal{N}, \forall{t} \in \mathcal{T} \label{eq:LL_net1}\\
	& && \delta_{i,\text{ref},t} = 0 \quad \forall {t} \in \mathcal{T} \label{eq:LL_net2}\\
	& && -P_{n,m}^{\text{lim}}\leq b_{n,m} \left ( \delta_{i,n,t} - \delta_{i,m,t}\right )\leq P_{n,m}^{\text{lim}} \notag\\
	& && \quad \forall n \in \mathcal{N}, m \in \Omega_{n}, \forall {t} \in \mathcal{T} \label{eq:LL_net3}\\
	& && 0 \leq P_{i,r} \leq P_r^\text{max} - \sum_{j \in \mathcal{I}, j < i} P_{j,r}, \quad \forall{r} \in \mathcal{R} \label{eq:LL_prb}\\
	& && 0 \leq P_{i,o} \leq \left ( \sum_{j \in \mathcal{I}, j \leq i} s_{j,o}^{\text{i}} \right) (P_o^\text{max} - \sum_{j \in \mathcal{I}, j < i} P_{j,o}),  \notag\\
	& && \quad \forall{o} \in \mathcal{O} \label{eq:LL_pob}\\
	& && P_{k}^{\text{i,U}} = AR_{i,k} s_{i,k} P_{k}^\text{U,max}, \quad \forall{k} \in \mathcal{K} \label{eq:LL_bbu}\\
	& && P_{k}^{\text{i,D}} = AR_{i,k} s_{i,k} P_{k}^\text{D,max}, \quad \forall{k} \in \mathcal{K} \label{eq:LL_bbd}\\
	& && AR_{i,k} \in \left\{0,1 \right\} \quad \forall k \in \mathcal{K} \label{eq:LL_AR_bin}
  \end{align}
  \endgroup
  \label{eq:problem2bis}
\end{subequations}
\normalsize
where $\mathbf{y}=\{P_{i,r},P_{i,o},AR_{i,k}, P_{i,k}^\text{U}, P_{i,k}^\text{D}, \delta_{i,n,t}\}$.
Whenever a new bid is submitted, the continuous market matches the bids in order to maximize the social welfare for the current match, which is given by the objective function \eqref{eq:LL_obj}. Equations \eqref{eq:LL_sumU} and \eqref{eq:LL_sumD} ensure that offers can only match with requests, in up and down directions, where $\mathcal{O}_{t}^{\text{U}}$, $\mathcal{O}_{t}^{\text{D}}$, $\mathcal{R}_{t}^{\text{U}}$, $\mathcal{R}_{t}^{\text{D}}$ and $\mathcal{K}_{t}$ are subsets of $\mathcal{O}$, $\mathcal{R}$ and $\mathcal{K}$ for upward (U) and downward (D) bids, submitted for time period $t$. The network constraints are given in \eqref{eq:LL_net1}, \eqref{eq:LL_net2} and \eqref{eq:LL_net3}, similarly as shown in the previous sections. The main difference here is that the initial setpoint is modified by all the offers and requests that have been accepted in previous rounds of the continuous market clearing. Constraints \eqref{eq:LL_prb} and \eqref{eq:LL_pob} give the bounds for each request and single offer. The maximum is equal to the quantity bid minus what has been previously accepted. The offers can only be accepted if they were previously submitted, which is why we need the sum over the binary variables here. Regarding the block bids, with \eqref{eq:LL_bbu}, \eqref{eq:LL_bbd} and \eqref{eq:LL_AR_bin}, the all-or-nothing condition is ensured.
Note that with this formulation, we do not consider that for two potential matches at the same price, the oldest bid would be prioritized. The formulation could be completed to include this, however it is not critical as we are mostly interested in determining the total resulting social welfare, which will not be impacted by this.

The resulting bilevel problem can be reformulated as a single level problem in order to be solved. To do this, it is necessary to relax the binary conditions \eqref{eq:LL_AR_bin}, as done in \cite{ye2019}. The binary conditions can be reintroduced in the resulting problem but the reformulation is then not exact. Solving the problem with the relaxed binary constraints gives a lower bound on the worst social welfare. It is exactly equal to the worst social welfare (i.e. the relaxation is tight) if the corresponding variables only take 1 and 0 values at the optimal point. Otherwise, it is possible to reintroduce the binary constraints and the exact worst social welfare will be somewhere between the two values obtained.

\subsection{Results for a Small Test-Case}
We solved this problem on a 5-bus test network, with 5 bids submitted on two time periods, including one block bid. The corresponding code and data used are available online \cite{github}. To reformulate the problem, the KKT conditions of the relaxed lower-level are used and the big-M method is applied to linearize the complementarity constraints. The solutions found for the relaxed problem are feasible for the original problem, which guarantees their optimality. For this system, we determine that the social welfare will be between 83\% and 100\% of the value obtained with auction-based clearing for the same set of bids.

The proposed algorithm is a computationally heavy problem, with the complexity increasing exponentially with the number of bids. The number of binaries and the number of lower-level problems to be solved are both equal to the squared number of offer bids considered. Future work will focus on reducing this complexity.

\section{Discussion}
\label{sec:discussion}

\subsection{Models Formulation}

Using auction-based or continuous clearing in LFMs has different implications besides the market outcome. The main distinction lies in their formulation. 

On one hand, the continuous matching algorithm is constantly comparing the incoming bids with the unmatched ones previously submitted, performing a network check every time there is a match and updating the network setpoint when a match is feasible. It also identifies and stores all candidate matches for the block bids and verifies the status of all the offers involved whenever one of them is matched. All of these processes increase the algorithm's complexity, but bids match as soon as possible. Flexibility requests and offers can be submitted close to real-time, which allows the latest forecasts to be used. 

On the other hand, the auction-based market is defined as an optimisation problem focused on minimising the total system costs. It is not straightforward to solve though, as (i) it is a MILP and (ii) it implies clearing all the time periods at the same time to accommodate the bids that link several periods. Thus, all the bids must be submitted before the market clearing, regardless of their time target. While continuous clearing can be performed in an online fashion, auction-based clearing requires the definition of a market time horizon. 

\subsection{Block Bids}

Block bids are mainly introduced in the LFM to model the rebound effect suffered by certain flexible loads. In order to handle technical requirements of flexibility providers they are subject to the AoN acceptance condition. This condition is also attached to block bids in most intraday markets for the sake of simplicity.

Block bids could also be attached to less restrictive conditions, such as being partially accepted or having a minimum acceptance ratio. Such kind of block bids could allow more flexibility trades and also model other providers' preferences. However, the introduction of different type of block bids would significantly increase the computational time of the market clearing, which is a very valuable property for markets operating close to real time. For this reason we decided to have only fully accepted or rejected block bids. Moreover, the other types of block bids mentioned can be simulated by submitting several smaller block bids or by combining single bids.

\section{Conclusion}
\label{sec:ccl}

This paper has two main contributions. First, it introduces a continuous market clearing algorithm for local flexibility markets which considers, for the first time to our knowledge, both network constraints and block bids. Continuous markets can allow a higher market liquidity, as flexibility offers and requests are matched continuously, and can result to larger trading volumes. The algorithm we propose in this paper focuses on energy flexibility markets in the distribution grid. Such markets have the potential to harness the available flexibility in the distribution system and defer costly network reinforcements. During the online matching, our proposed algorithm considers the network constraints to guarantee the technical feasibility of the trades and avoid causing or aggravating congestions in the distribution grid.
At the same time, it enables block bids, which increases the pool of available flexibility as it promotes the participation of potential flexibility suppliers that suffer a rebound effect when providing the service.

The second main contribution of this paper is the design of an algorithm that determines the upper and lower bound of the suboptimality of a continuous energy market compared with its auction-based counterpart. Compared to an auction-based clearing, which considers all flexibility offers and requests for the total time horizon at the same time, continuous markets result by design to a lower social welfare; this is often seen as an acceptable shortcoming by regulators and market operators, when considering the increased liquidity that continuous markets often result to, and their ability to operate close to real-time. This paper formulates an optimization algorithm that can determine the worst and best arrival sequence of the bids in a continuous market, and through that, the best and worst performance in terms of social welfare, compared to an auction-based market. Considering a known set of possible bids, our results on a 5-bus system show that the continuous market (which includes both network constraints and block bids) will result to a social welfare between 83\% and 100\% of the social welfare that an auction-based market can achieve. 

Besides the design of an algorithm, we also carried out a series of simulations on a larger system to investigate the difference in social welfare and traded volume between the two market clearing models for a series of different conditions. Running our model for 100 different bid arrival sequences, we find that the average social welfare is at 88.6\% of the optimal. Regarding liquidity, we find that with single bids it is very likely to end up with a higher energy volume traded through continuous clearing. This is not the case, however, when we allow the trade of block bids, where the average energy volume traded is 89.5\% lower than with auction-based clearing.

\bibliographystyle{IEEEtran}
\bibliography{library.bib}

\end{document}